\documentclass[fleqn,12pt,twoside]{article}
\usepackage[headings]{espcrc1}
\usepackage{amsmath}
\usepackage{graphicx}
\usepackage{cite}

\newcommand{\ds}{\displaystyle}

\def\eqref#1{(\ref{#1})}
\def\l{\langle}

\def\r{\rangle}

\def\o{{\cal O}}
\def\const{\mathrm{const}}

\title{Generalized-ensemble simulations and cluster algorithms}

\author{M. Weigel\address{Theoretische Physik, Universit\"at des Saarlandes, D-66041 Saarbr\"ucken, Germany}%
  \address{Institut f\"ur Physik, Johannes Gutenberg-Universit\"at Mainz,
    Staudinger Weg 7, D-55099 Mainz, Germany}\thanks{Email: weigel@uni-mainz.de},
}

\runtitle{Generalized ensembles and clusters}
\runauthor{M. Weigel}

\begin{document}

\maketitle

\begin{abstract}
  The importance-sampling Monte Carlo algorithm appears to be the universally optimal
  solution to the problem of sampling the state space of statistical mechanical
  systems according to the relative importance of configurations for the partition
  function or thermal averages of interest. While this is true in terms of its
  simplicity and universal applicability, the resulting approach suffers from the
  presence of temporal correlations of successive samples naturally implied by the
  Markov chain underlying the importance-sampling simulation. In many situations,
  these autocorrelations are moderate and can be easily accounted for by an
  appropriately adapted analysis of simulation data. They turn out to be a major
  hurdle, however, in the vicinity of phase transitions or for systems with complex
  free-energy landscapes. The critical slowing down close to continuous transitions
  is most efficiently reduced by the application of cluster algorithms, where they
  are available. For first-order transitions and disordered systems, on the other
  hand, macroscopic energy barriers need to be overcome to prevent dynamic ergodicity
  breaking. In this situation, generalized-ensemble techniques such as the
  multicanonical simulation method can effect impressive speedups, allowing to sample
  the full free-energy landscape. The Potts model features continuous as well as
  first-order phase transitions and is thus a prototypic example for studying phase
  transitions and new algorithmic approaches. I discuss the possibilities of bringing
  together cluster and generalized-ensemble methods to combine the benefits of both
  techniques. The resulting algorithm allows for the efficient estimation of the
  random-cluster partition function encoding the information of all Potts models,
  even with a non-integer number of states, for all temperatures in a single
  simulation run per system size.
\end{abstract}

\section{Introduction}

The Monte Carlo simulation method has developed into one of the standard tools for
the investigation of statistical mechanical systems undergoing first-order or
continuous phase transitions \cite{binder:book2}. While the formulation of the
Metropolis-Hastings algorithm \cite{metropolis:53a,hastings:70}, which is the basic
workhorse of the method up to this very day, dates back more than half a century ago,
its initial practical value was limited. This was partially due to the the fact that
computers for the implementation of such simulations where not widely available and
the computing power of those at hand was very limited compared to today's
standards. Hence, the method was not yet competitive, e.g., for studying critical
phenomena compared to more traditional approaches such as the $\epsilon$ or series
expansions \cite{zinn-justin}. That the situation has changed rather drastically in
favor of the numerical approaches since those times is not only owed to the dramatic
increase in available computational resources, but probably even more importantly to
the invention and refinement of a number of advanced techniques of simulation and
data analysis. These include (but are not limited to) the introduction of the concept
of finite-size scaling \cite{barber:domb}, which turned the apparent drawback of
finite system sizes in simulations into a powerful tool for extracting the asymptotic
scaling behavior, the invention of cluster algorithms \cite{swendsen:86,wolff:89a}
beating the critical slowing down close to continuous transitions, the
(re-)introduction of histogram reweighting methods
\cite{ferrenberg:88a,ferrenberg:89a} allowing for the extraction of a continuous
family of estimators of thermal averages from a single simulation run, and the
utilization of a growing family of generalized-ensemble simulation techniques such as
the multicanonical method \cite{berg:92b}, that allow to overcome barriers in the
free-energy landscape and enable us to probe highly suppressed transition states.

\begin{figure}[t]
  \centering
  \scalebox{0.9}{
    \input{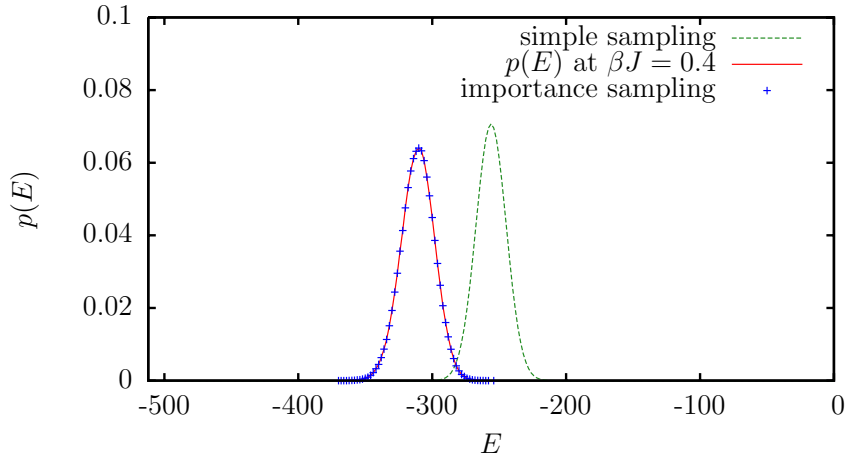}
  }
  \vspace*{-0.5cm}
  \caption
  { Probability distributions of reduced energies $E\equiv {\cal H}/J$ for the
    example of a $16\times 16$ $2$-state Potts model at coupling $\beta J = 0.4$,
    cf.~Eq.~(\ref{eq:potts_hamiltonian}). For the case of simple sampling, the
    overlap of $p_\mathrm{sim}(E)$ and $p_\mathrm{eq}(E)$ is vanishingly small,
    whereas the two distributions coincide exactly for the case of importance
    sampling.
    \label{fig:importance_sampling}
  }
\end{figure}

In a general setup for a Monte Carlo simulation, the microscopic states of the system
appear with frequencies according to a probability distribution
$p_\mathrm{sim}(\{s_i\})$, which is an expression of the chosen prescription of
picking states and hence specific to the used simulation algorithm. Here, having a
spin system in mind, we label the states with the set $\{s_i\}$, $i=1,\ldots,V$, of
variables. In thermal equilibrium, on the other hand, microscopic states are
populated according to the Boltzmann distribution for the case of the canonical
ensemble,
\begin{equation}
  \label{eq:canonical_distr}
  p_\mathrm{eq}(\{s_i\}) = \frac{1}{Z_\beta}e^{-\beta{\cal H}(\{s_i\})},
\end{equation}
where ${\cal H}(\{s_i\})$ denotes the energy of the configuration $\{s_i\}$ and
$Z_\beta$ is the partition function at inverse temperature $\beta = 1/T$. Therefore,
any sampling prescription with non-zero probabilities for all possible microscopic
states $\{s_i\}$ in principle\footnote{If the samples are generated by a Markov chain
  there are, of course, additional caveats. In particular, any two states must be
  connected by a finite sequence of transitions of positive probability, i.e., the
  Markov chain must be ergodic.} allows to estimate thermal averages of any
observable ${\cal O}(\{s_i\})$ from a time series $\{s_i^{(t)}\}$, $t=1,\ldots,N$, of
observations,
\begin{equation}
  \label{eq:general_estimate}
  \hat{O} = \frac{\ds\sum_{t=1}^N {\cal O}(\{s_i^{(t)}\})\frac{\ds
      p_\mathrm{eq}(\{s_i^{(t)}\})}{\ds p_\mathrm{sim}(\{s_i^{(t)}\})}}
  {\ds\sum_{t=1}^N \frac{\ds p_\mathrm{eq}(\{s_i^{(t)}\})}{\ds p_\mathrm{sim}(\{s_i^{(t)}\})}},
\end{equation}
such that $O\equiv\l {\cal O}\r = \l \hat{O}\r$ at least in the limit
$N\rightarrow\infty$ of an infinite observation time. For a finite number of samples,
however, the estimate (\ref{eq:general_estimate}) becomes unstable as soon as the
simulated and intended probability distributions are too dissimilar, such that only a
vanishing number of simulated events are representative of the equilibrium
distribution. This is illustrated in Fig.~\ref{fig:importance_sampling}, where the
distribution $p_\mathrm{sim}(\{s_i\}) = \const.$ of purely random or simple sampling
is compared to the canonical distribution (\ref{eq:canonical_distr}) for a finite
temperature $T$. Since the Boltzmann distribution (\ref{eq:canonical_distr}) only
depends on the energy $E = {\cal H}(\{s_i\})$, it is here useful to compare the
one-dimensional densities $p_\mathrm{sim}(E)$ and $p_\mathrm{eq}(E)$ instead of the
high-dimensional distributions in state space. For importance sampling, on the other
hand, the simulated probability density is identical to the equilibrium distribution
which can be achieved, e.g., by proposing local updates $s_i\rightarrow s_i'$ (spin
flips) at random and accepting them according to the Metropolis rule
\begin{equation}
  \label{eq:metropolis}
  T(\{s'_i\}|\{s_i\})=\min[1,p_\mathrm{eq}(\{s'_i\})/p_\mathrm{eq}(\{s_i\})]
\end{equation}
for the transition probability $T(\{s'_i\}|\{s_i\})$.

While importance sampling is optimal in that the simulated and intended probability
densities coincide, this benefit comes at the expense of introducing correlations
between successive samples. Under these circumstances, the autocorrelation function
of an observable ${\cal O}$ is expected to decay exponentially,
\begin{equation}
  \label{eq:tau_def}
  C_{\cal O}(t) \equiv \l \o_0 \o_t \r -\l \o_0 \r\l \o_t\r \sim C_\o(0) e^{-t/\tau(\o)},  
\end{equation}
defining the associated {\em autocorrelation time\/} $\tau({\cal
  O})$. Autocorrelations are a direct limiting factor for the amount of information
that be extracted from a time series of a given length for estimating thermal
averages. This is most clearly seen by inspecting an alternative definition of
autocorrelation time involving an integral or sum of the autocorrelation function,
\begin{equation}
  \tau_{{\rm int}}(\o) \equiv
  \frac{1}{2}+\sum_{t=1}^{\infty}C_{\cal O}(t)/C_{\cal O}(0).
  \label{tau_int_def}
\end{equation}
The resulting {\em integrated autocorrelation time\/} determines the precision of an
estimate $\hat{O}$ for the thermal average $\l\o\r$ from the time series
\cite{sokal:92a},
\begin{equation}
  \label{eq:precision_autocorrelations}
  \sigma^2(\hat{O}) \approx \frac{\sigma^2(\o)}{N/2\tau_\mathrm{int}(\o)}.
\end{equation}
The presence of autocorrelations hence effectively reduces the number of independent
measurements by a factor $1/2\tau_\mathrm{int}(\o)$. Generically, autocorrelation
times are moderate and the problem is thus easily circumvented by adapting the number
of measurement sweeps according to the value of $\tau_\mathrm{int}$. The problem
turns out to be much more severe, however, in the vicinity of phase transitions
points. Close to a critical point, where clusters of pure phase states of all sizes
constitute the typical configurations, one observes {\em critical slowing
  down\/}\footnote{The exponential autocorrelation time $\tau$ of
  Eq.~(\ref{eq:tau_def}) and the integrated autocorrelation time $\tau_\mathrm{int}$
  of Eq.~(\ref{tau_int_def}) are found to exhibit the same asymptotic scaling
  behavior, such that we do not distinguish them in this respect.},
\begin{equation}
  \tau \sim \min(\xi,L)^z,
  \label{dyn_exp}
\end{equation}
where the {\em dynamical critical exponent\/} $z$ is found to be close to $z=2$ for
conventional local updating moves. Since the correlation length $\xi$ diverges as the
transition is approached, the same holds true for the autocorrelation time. This
problem is most elegantly solved by the introduction of cluster algorithms, which
involve updating collective variables that happen to show incipient percolation right
at the ordering transition of the spin system and, in addition, must exhibit
geometrical properties which are commensurate with the intrinsic geometry of the
underlying critical system. Such approaches are discussed for the case of the Potts
model in the context of the random-cluster representation in Section
\ref{sec:random_cluster} below.

\begin{figure}[t]
  \centering
  \scalebox{0.9}{
    \input{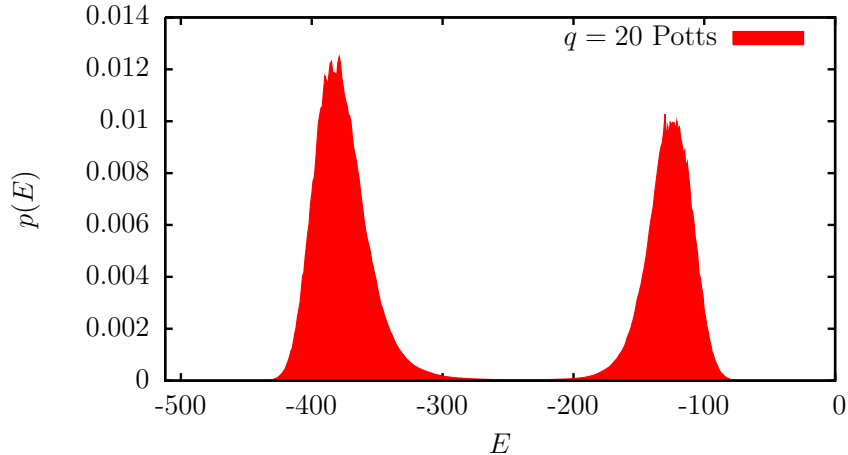}
  }
  \vspace*{-0.5cm}
  \caption
  { Sampled probability distribution of the internal energy $E$ of the $q=20$ states
    Potts model on a $16\times 16$ square lattice at the transition coupling
    $\beta J=1.699669\ldots$
    \label{fig:first_order}
  }
\end{figure}

Even more dramatic correlation effects are seen for the case of first-order phase
transitions. There, transition states connecting the pure phases coexisting at the
transition are highly suppressed, leading to the phenomenon of {\em metastability\/},
where the phase of higher free energy persists in some region below the transition
point for macroscopic times due to the free-energy barrier that needs to be overcome
to effect the ordering of the system. This is illustrated in
Fig.~\ref{fig:first_order} displaying the order parameter distribution of a $q=20$
states Potts model. The mixed phase states connecting the peaks of the pure phases
correspond to configurations containing interfaces and consequently carry an extra
excitation energy $\sim 2\sigma L^{d-1}$, where $L$ denotes the linear size
of the system and $\sigma$ is the surface free-energy per unit area associated to
interfaces between the pure phases. The thermally activated dynamics in overcoming
this additional energy barrier leads to exponentially divergent autocorrelation
times,
$$
\tau \sim \exp(-2\beta\sigma L^{d-1}),
$$
sometimes referred to as hypercritical slowing down. Due to the finite correlation
length at the transition point, cluster updates are of no use here but, instead,
techniques for overcoming energy barriers are required. These can be provided
(besides other means) by generalized-ensemble simulations discussed below in Section
\ref{sec:generalized_ensembles}. Similar problems are encountered in simulations of
disordered systems, where a multitude of metastable states separated by barriers is
found \cite{holovatch:07}.

The Potts model is a natural extension of the Ising model of a ferromagnet to a
system with $q$-state spins and Hamiltonian
\begin{equation}
  \label{eq:potts_hamiltonian}
    {\cal H} = -J \sum_{\langle i,j\rangle}\delta(\sigma_i,\sigma_j),
  \;\;\;\;\sigma_i = 1,\ldots,q.
\end{equation}
It is well known that the Potts model undergoes a continuous phase transitions for
$q\le 4$ in two dimensions and $q < 3$ in three dimensions, while the transition
becomes discontinuous for a larger number of states \cite{wu:82a}. The $q=2$ Potts
model is equivalent to the well-known Ising model. In the random-cluster
representation introduced below in Section \ref{sec:random_cluster}, the definition
of the Potts model is naturally extended to all real values of $q > 0$, and it turns
out that the (bond) percolation problem then corresponds to the limit $q\rightarrow
1$, while for $q\rightarrow 0$ the model describes random resistor networks. Due to
these properties, the Potts model serves as a versatile playground for the study of
phase transitions with applications ranging from condensed matter to high-energy
physics.

\section{Using histograms}

When studying phase transitions with Markov chain Monte Carlo simulations, one
encounters another generic problem independent of the presence of
autocorrelations. In the standard approach, estimators of the type
(\ref{eq:general_estimate}) need to be used for each of a series of independent
simulations at different values of the temperature (or other external parameters) to
extract the temperature dependence of the observable at hand. This turns out to be
problematic when studying phase transitions, where certain observables (such as,
e.g., the specific heat) develop peaks which are narrowing down to the location of
the transitions point as successively larger system sizes are considered. Locating
such maxima to high precision then requires to perform a large number of independent
simulation runs.

\subsection{Energy and magnetization histograms}

\begin{figure}[t]
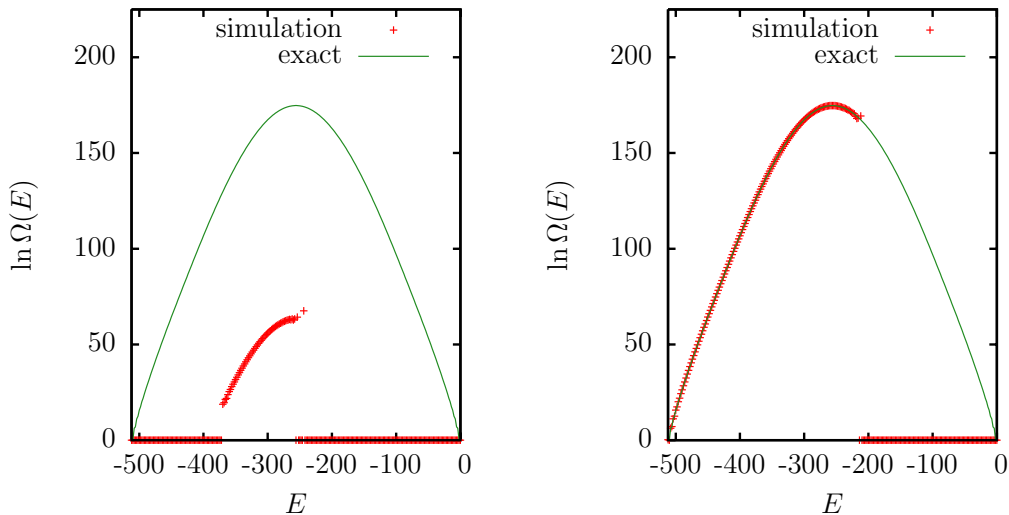

  \centering
  \scalebox{0.9}{
    \input{q2-0c4-densityT}
  }
  \scalebox{0.9}{
    \input{q2-density-normalizedT}
  }
  \vspace*{-0.5cm}
  \caption
  { Density-of-states estimate $\hat{\Omega}(E)$ for the $2$-state Potts model on a
    $16\times 16$ square lattice from importance-sampling simulations at coupling
    $K=0.4$ (left panel) and from a series of simulations ranging from $K=0.1$ up to
    $K=0.9$ (right panel). The solid lines show the exact density of states
    calculated according to Ref.~\cite{beale:96a}. From the estimate
    (\ref{eq:single_histogram_dos}) and the corresponding multi-histogram analogue
    (\ref{eq:optimal_edos}), $\Omega(E)$ can only be determined up to an unknown
    normalization.
    \label{fig:beale}
  }
\end{figure}

This problem is avoided by realizing that each time series from a simulation run at
fixed temperature can be used to estimate thermal averages for nearby temperatures as
well. The concept of {\em histogram reweighting\/}
\cite{ferrenberg:88a,ferrenberg:89a} follows directly from the general relation
(\ref{eq:general_estimate}) connecting simulated and target probability densities. If
an importance sampling simulation is performed at coupling $K_0 = \beta_0 J$,
i.e., $p_\mathrm{sim}\sim \exp(-K_0 E)$, estimators for canonical expectation
values at a different coupling $K$ are found from Eq.~(\ref{eq:general_estimate})
with $p_\mathrm{eq}\sim\exp(-K E)$,
\begin{equation}
   \hat{O}_K = \frac{\sum_{t=1}^N {\cal
       O}(\{s_i^{(t)}\})e^{-(K-K_0){\cal H}(\{s_i^{(t)}\})/J}}
  {\sum_{t=1}^Ne^{-(K-K_0){\cal H}(\{s_i^{(t)}\})/J} }.
  \label{eq:basic_reweighting}
\end{equation}
While this is conceptually perfectly general, it is clear that --- quite similar to
the case of simple sampling discussed above --- reliable estimates can only be
produced if the simulated and target distributions have significant overlap (cf.\
Figure \ref{fig:importance_sampling}). This is most clearly seen when switching over
to a formulation involving histograms as estimates of the considered probability
densities. If $\hat{H}_{K_0}(E)$ is a sampled energy histogram at coupling $K_0$ and
the observable ${\cal O}$ only depends on the configuration $\{s_i\}$ via the energy
$E$, the estimate (\ref{eq:basic_reweighting}) becomes
\begin{equation}
  \hat{O}_K = \frac{\sum_E{\cal O}(E)\hat{H}_{K_0}(E)e^{-(K-K_0)E}}
  {\sum_E\hat{H}_{K_0}(E)e^{-(K-K_0)E}}.
  \label{eq:histogram_reweight}
\end{equation}
It is useful to realize, then, that sampling the histogram $\hat{H}_{K_0}(E)$ one is,
in fact, estimating the {\em density of states\/} $\Omega(E)$,
\begin{equation}
  \langle \hat{H}_{K_0}(E)/N \rangle = p_{K_0}(E) = \frac{1}{Z_{K_0}}\Omega(E)e^{-K_0E}
  \label{eq:helper17}
\end{equation}
i.e., the number of microstates of energy $E$ via
\begin{equation}
  \hat{\Omega}(E) = Z_{K_0}\,\hat{H}_{K_0}(E)/N\times e^{K_0 E},
  \label{eq:single_histogram_dos}
\end{equation}
where $Z_{K_0}$ denotes the partition function at coupling $K_0$. Inserting this
expression into Eq.~(\ref{eq:helper17}), one indeed arrives back at the reweighting
estimate (\ref{eq:histogram_reweight}),
\begin{equation}
  \hat{O}_K = \frac{1}{Z_K} \sum_E \hat{\Omega}(E) e^{-KE} {\cal O}(E) = \frac{Z_{K_0}}{Z_K}\sum_E
  \hat{H}_{K_0}(E)/N \times e^{-(K-K_0)E}{\cal O}(E)
\end{equation}
with
\begin{equation}
  \widehat{\frac{Z_{K_0}}{Z_K}} =
  \frac{\sum_E\hat{H}_{K_0}(E)/N\times e^{-(K_0-K_0)E}}{\sum_E\hat{H}_{K_0}(E)/N\times e^{-(K-K_0)E}}.
\end{equation}
It should be clear that the density of states is a rather universal quantity in that
its complete knowledge allows to determine any thermal average related to the energy
for arbitrary temperatures. The limitation in the allowable reweighting range,
$|K-K_0|<\epsilon$, then translates into a window of energies for which the density
of states $\Omega(E)$ can be reliably estimated from a single canonical
simulation. This is illustrated in the left panel of Figure \ref{fig:beale} for the
$2$-state Potts model, where the density-of-states estimate of
Eq.~(\ref{eq:single_histogram_dos}) from a single simulation at coupling $K=0.4$ is
compared to the exact result. Note that from the estimator
(\ref{eq:single_histogram_dos}) $\Omega(E)$ can only be determined up to the unknown
normalization constant $Z_{K_0}$. This is irrelevant for thermal averages of the type
(\ref{eq:histogram_reweight}), but precludes the determination of free energies.

If more than a tiny patch of the domain of the density of states is to be determined,
several simulations at different couplings $K$ need to be combined. Since the average
internal energy increases monotonically with temperature, a systematic series of
simulations can cover the relevant range of energies. A combination of several
estimates of the form (\ref{eq:single_histogram_dos}) for $\Omega(E)$ is problematic,
however, since each estimate has a different, unknown scale factor $Z_{K_i} = e^{\cal
  F}_i$. This dilemma can be solved by the iterative solution of a system of
equations. A convex linear combination of density-of-states estimates
$\hat{\Omega}_i(E)$ from independent simulations at couplings $K_i$, $i=1,\ldots,n$,
$$
\hat{\Omega}(E) \equiv \sum_{i=1}^n\alpha_i(E)\hat{\Omega}_i(E),
$$
with $\sum_i\alpha_i(E) = 1$ is of minimal variance for \cite{weigel:09}
$$
\alpha_i(E) = \frac{1/\sigma^2[\hat{\Omega}_i(E)]}{\sum_i 1/\sigma^2[\hat{\Omega}_i(E)]}.
$$
In view of Eq.~(\ref{eq:single_histogram_dos}), and ignoring the variance of the
scale factors $e^{{\cal F}_i}$, one estimates
$$
\sigma^2[\hat{\Omega}_i(E)] \approx e^{2{\cal F}_i}\hat{H}_{K_i}(E)/N^2\times e^{2K_i E},
$$
such that
\begin{equation}
  \hat{\Omega}(E) = \frac{\sum_ie^{-{\cal F}_i-K_i E}}{\sum_i e^{-2{\cal F}_i-2K_iE}[\hat{H}_{K_i}(E)/N]^{-1}}.
  \label{eq:optimal_edos}
\end{equation}
From Eq.~(\ref{eq:single_histogram_dos}) follows the normalization condition
\begin{equation}
  \label{eq:iteration2}
  e^{\cal F}_i = \sum_E\hat{\Omega}(E) e^{-K_i E},
\end{equation}
which needs to be solved iteratively with Eq.~(\ref{eq:optimal_edos}) to
simultaneously result in the optimized estimate $\hat{\Omega}(E)$ and the scale
factors ${\cal F}_i$. An initial estimate can be deduced from thermodynamic
integration \cite{ferrenberg:89a,weigel:02a,binder:book2}. Here, again, it is a
crucial condition that the energy histograms to be combined have sufficient
overlap. Otherwise, the iterative solution of Eqs.~(\ref{eq:optimal_edos}) and
(\ref{eq:iteration2}) cannot converge. Combining an appropriately chosen series of
simulations, from this {\em multi-histogram\/} approach a reliable estimate of the
full density of states can be achieved, as is illustrated in the right panel of
Figure \ref{fig:beale} for the case of the $2$-state Potts model. (The states to the
right of the maximum in $\Omega(E)$ belong to the {\em antiferromagnetic\/} Potts
model and thus are not seen in the simulations.) If the full range of admissible
energies has been sampled, also an {\em absolute\/} normalization of $\Omega(E)$
becomes possible, matching $\hat{\Omega}(E)$ to reproduce the number $q$ of ground
states or the number $q^{\cal N}$ of different states in total, where ${\cal N}$
denotes the number of Potts spins.

For estimating thermal averages of observables that do not depend on the energy only,
the outlined framework can be easily generalized by replacing the measurements ${\cal
  O}(E)$ in Eq.~(\ref{eq:histogram_reweight}) by the corresponding microcanonical
averages $\l{\cal O}\r_E$ at energy $E$,
$$
\hat{O}_K = \frac{\sum_E\l{\cal O}\r_E\hat{H}_{K_0}(E)e^{-(K-K_0)E}}
  {\sum_E\hat{H}_{K_0}(E)e^{-(K-K_0)E}}.
$$
In the context of spin models, for instance, it can be useful to sample joint
histograms of energy and magnetization and also define the corresponding
two-dimensional density of states \cite{tsai:unpublished}. For the Potts model,
however, it appears to be even more natural to consider a density of states occurring
in the random cluster representation which also is the natural language for the
formulation of cluster algorithms. This will be discussed in the next section.

\subsection{Random cluster histograms\label{sec:random_cluster}}

As was first noted by Fortuin and Kasteleyn \cite{fortuin:72a}, the partition
function of the zero-field Potts model on a general graph ${\cal G}$ with ${\cal N}$
vertices and ${\cal E}$ edges can be rewritten as
\begin{equation}
  Z_{K,q} \equiv \sum_{\{\sigma_i\}} e^{K\sum_{\langle i,j\rangle}\delta(\sigma_i,\sigma_j)}
  = \sum_{{\cal G}' \subseteq {\cal G}} (e^K-1)^{b({\cal G}')}\,q^{n({\cal G}')},
  \label{eq:random_cluster}
\end{equation}
where the sum runs over all bond configurations ${\cal G}'$ on the graph
(subgraphs). Note that the formulation (\ref{eq:random_cluster}) in contrast to that
of Eq.~(\ref{eq:potts_hamiltonian}) allows for a natural continuation of the model to
{\em non-integer\/} values of $q$.  This expression can be interpreted as a
bond-correlated percolation model with percolation probability $p = 1-e^{-K}$:
\begin{equation}
  Z_{p,q} = e^{K{\cal E}} \sum_{{\cal G}' \subseteq {\cal G}}
  p^{b({\cal G}')}(1-p)^{{\cal E}-b({\cal G}')}\,q^{n({\cal G}')}
  = e^{K{\cal E}} \sum_{b=0}^{{\cal E}}
  \sum_{n=1}^{{\cal N}}g(b,n)\,p^b\,(1-p)^{{\cal E}-b}\,q^n,
  \label{eq:partition_bcpm}
\end{equation}
where $g(b,n)$ denotes the number of subgraphs of ${\cal G}$ with $b$ activated bonds
and $n$ resulting clusters. This purely combinatorial quantity corresponds to the
density of states of the random-cluster model.

It is this formulation of the model which is exploited by the cluster algorithms
\cite{swendsen:86,wolff:89a} mentioned above. Since the Potts model is {\em
  equivalent\/} to a (correlated) percolation model, it follows (almost)
automatically that the thus defined clusters percolate at the ordering transition and
have the necessary fractal properties. This deep connection between spin model and
percolation problem results in cluster algorithms for the Potts model dramatically
reducing, and in some cases completely removing, the effect of critical slowing down
\cite{wj:chem}. It appears thus desirable to combine this extraordinarily successful
approach with the idea of reweighting to result in continuous families of
estimates. In particular, one would want to reweight in the temperature as well as
the now continuous parameter $q$, for instance for determining the tricritical value
$q_c$ where the transition becomes of first order. In contrast to previous attempts
in this direction \cite{lee:91a} using the language of energy and magnetization that
results in certain systematic errors, such reweighting is very naturally possible in
the random-cluster representation. By construction, a cluster-update simulation of
the $q_0$-state Potts model at coupling $K_0$ produces bond configurations with the
probability distribution
\begin{equation}
  p_{p_0,q_0}(b,n) = W_{p_0,q_0}^{-1}\,g(b,n)\,p_0^b\,(1-p_0)^{{\cal
      E}-b}\,q_0^n,
  \label{eq:rc_probability}
\end{equation}
where $p_0=1-e^{-K_0}$ and $W_{p_0,q_0} \equiv Z_{p_0,q_0} e^{-K_0{\cal
    E}}$. Therefore, if a histogram $\hat{H}_{p_0,q_0}(b,n)$ of bond and cluster
numbers is sampled, one has $p_{p_0,q_0}(b,n) =
\langle\hat{H}_{p_0,q_0}(b,n)/N\rangle$ and thus follows an estimate of $g(b,n)$ as
\cite{weigel:02a}
\begin{equation}
  \hat{g}(b,n) = W_{p_0,q_0} \frac{\hat{H}_{p_0,q_0}(b,n)}{p_0^b\,(1-p_0)^{{\cal
        E}-b}\,q_0^n\,N},
\end{equation}
which, analogous to the estimate (\ref{eq:single_histogram_dos}), contains an
(unknown) normalization factor, $W_{p_0,q_0}$. The required cluster decomposition of
the lattice is a by-product of the Swendsen-Wang update and hence its determination
does not entail any additional computational effort.

In this way, cluster-update simulations with largely reduced critical slowing down
can be used for a systematic study of the model for arbitrary temperatures and
(non-integer) numbers of states. Thermal averages of observables ${\cal O}(b,n)$
can be easily deduced from $\hat{g}(b,n)$,
\begin{equation}
  \hat{\cal O}(p,q) \equiv \left[{\cal O}\right]_{p,q} = \frac{\ds\sum_{b=0}^{{\cal E}} \sum_{n=1}^{{\cal N}}
    \hat{H}_{p_0,q_0}(b,n)
    \left(\frac{p}{p_0}\right)^b\left(\frac{1-p}{1-p_0}\right)^{{\cal
        E}-b}\left(\frac{q}{q_0}\right)^n {\cal O}(b,n)}{\ds\sum_{b=0}^{{\cal E}} \sum_{n=1}^{{\cal N}}
    \hat{H}_{p_0,q_0}(b,n)
    \left(\frac{p}{p_0}\right)^b\left(\frac{1-p}{1-p_0}\right)^{{\cal
        E}-b}\left(\frac{q}{q_0}\right)^n}.
\end{equation}
Relating expressions in the $(b,n)$ and $(E,M)$ languages, we have,
\begin{eqnarray}
  \hat{u} & = & -\frac{1}{p{\cal N}}\,[b]_{p,q}, \nonumber \\
  \hat{c}_v & = & \frac{K^2}{p^2{\cal N}}\,\left(\left[(b-[b]_{p,q})^2\right]_{p,q}
    -(1-p)[b]_{p,q}\right), \nonumber
\end{eqnarray}
where $u$ denotes the internal energy per spin and $c_v$ is the specific heat. For
magnetic observables, an additional distinction between percolating and finite
clusters is necessary \cite{weigel:02a}.

\begin{figure}[t]
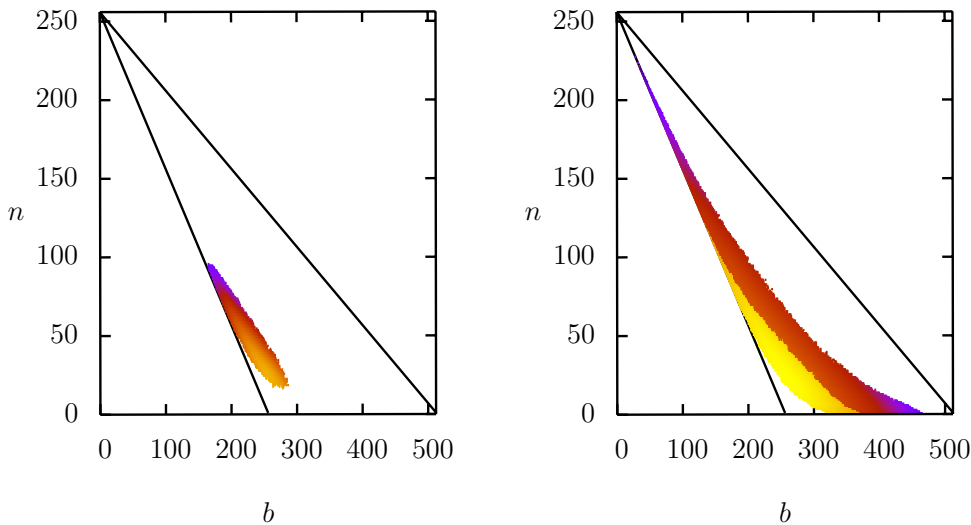

  \centering
  \vspace*{-0.8cm}
  \hspace*{0.3cm}
  \scalebox{0.9}{
    \input{support1T}
  }
  \scalebox{0.9}{
    \input{support2T}
  }
  \vspace*{-0.5cm}
  \caption
  { Random-cluster density of states $g(b,n)$ on the $16\times 16$ square lattice as
    estimated from a Swendsen-Wang cluster-update simulation of the $q=2$ Potts model
    at $K=0.8$ (left panel) and the $q=2$ (brighter part, below) as well as $q=10$
    (darker part, on top) models at a range of different couplings (right
    panel). Brighter colors correspond to larger values of $\hat{g}(b,n)$. The white
    areas correspond to $(b,n)$ values not visited in the simulations.
    \label{fig:support}
  }
\end{figure}

For averages at general values of $p$ and $q$, we run into the by now familiar
problem of vanishing overlap of histograms as we move too far from the simulated
$(p_0, q_0)$ point. This is illustrated in Figure \ref{fig:support}, showing the
support of the density-of-states estimate $\hat{g}$ in the $(b,n)$ plane. For a
single canonical simulation, only a small patch of the $(b,n)$ plane is sampled (left
panel). To improve on this, a multi-histogramming approach analogous to the technique
discussed in the previous section is required. The relations corresponding to
Eqs.~(\ref{eq:optimal_edos}) and (\ref{eq:optimal_edos}) are here \cite{weigel:02a}
\begin{equation}
  \hat{g}(b,n) = \frac{\sum_i N_i\,e^{-{\cal F}_i}\,p_i^b\,(1-p_i)^{{\cal
        E}-b}\,q_i^n}
  {\sum_j N_j^2\,e^{-2{\cal F}_j}\,p_j^{2b}\,(1-p_j)^{2({\cal E}-b)}\,
    q_j^{2n}[\hat{H}_{p_i,q_i}(b,n)]^{-1}},
  \label{selfcon_1}
\end{equation}
and the following self-consistency equation for the free-energy factors ${\cal F}_i$:
\begin{equation}
  e^{{\cal F}_i} = \sum_{b=0}^{\cal E}\sum_{n=1}^{\cal N}
  \hat{g}(b,n)\,p_i^b\,(1-p_i)^{{\cal E}-b}\,q_i^n.
  \label{selfcon_2}
\end{equation}
Combining a number of simulations at different temperatures and $q$ values, a more
significant patch of the density of states $g(b,n)$ can thus be sampled, cf.\ the
right panel of Figure \ref{fig:support}. Note that in the $(b,n)$ plane, moving from
$b=0$ to $b={\cal E}$ corresponds to moving from infinite to zero temperature,
whereas increasing the number of states $q$ moves the histograms up along the $n$
axis, corresponding to the fact that the presence of more states will tend to produce
configurations broken down into smaller (and thus more) clusters.

\section{Generalized ensembles\label{sec:generalized_ensembles}}

Two problems arise for an estimate of the total random-cluster density of states
$g(b,n)$ with the multi-histogram approach outlined above: (i) while simulations at
sufficiently small $q$ profit from the application of cluster algorithms in that
critical slowing down is strongly reduced, in the first-order regime of large $q$
cluster algorithms are not useful for tackling the hypercritical slowing down
observed there and (ii) as the system size is increased, histograms from simulations
at different (integer) values of $q$ cease to overlap, such that the set
(\ref{selfcon_1}) and (\ref{selfcon_2}) of multi-histogram equations eventually
breaks down. While the second problem could, in principle, be avoided by using the
cluster algorithm suggested in Ref.~\cite{chayes:98a} for general, non-integer $q$
values, we find it more convenient to tackle both issues simultaneously by moving
away entirely from the concept of {\em canonical\/} simulations which, as it turns
out, entails further advantages for the sampling problem.

The idea of multicanonical \cite{berg:92b} (or, less specifically, generalized
ensemble) simulations is motivated by the problem of dynamically tunneling the area
of (exponentially) low probability in between the coexisting phases at a first
order transition, cf.\ Figure \ref{fig:first_order}. Instead of simulating the
canonical distribution (\ref{eq:canonical_distr}), consider importance sampling
according to a generalized probability density,
\begin{equation}
  \label{eq:muca}
  p_\mathrm{muca}(E) = \frac{\Omega(E)/W(E)}{Z_\mathrm{muca}} =
  \frac{e^{S(E)-\omega(E)}}{Z_\mathrm{muca}},
\end{equation}
where $W(E)$ resp.\ $\omega(E)$ denote (logarithms of) suitably chosen weight factors
and $S(E) = \ln\Omega(E)$ is the microcanonical entropy. To overcome barriers, the
sampling distribution should be {\em broadened\/} with respect to the canonical one,
in the extremal case to become completely flat, $p_\mathrm{muca}(E) = \const$. For
this case Eq.~(\ref{eq:muca}) tells us that
$$
W(E) = \Omega(E)\;\;\;\text{resp.}\;\;\;\omega(E) = S(E).
$$
Hence, we arrive back at the task of estimating the density of states of the system!
Since $\Omega(E)$ is not known {\em a priori\/}, one needs to revert to a recursive
solution, where starting out, e.g., with the initial guess $W_0(E) = 1$
(corresponding to a canonical simulation at infinite temperature) one produces an
estimate $\hat{\Omega}_1(E)$ of the density of states according to
Eq.~(\ref{eq:single_histogram_dos}) and sets $W_1(E) = \hat{\Omega}_1(E)$. Repeating
this process, eventually a reliable estimate for $\Omega(E)$ over the full range of
energies can be produced\footnote{In practice it is, of course, more reasonable to
  combine the information from {\em all\/} previous simulations to form the current
  best guess for the weight function \cite{berg:96}.}. Note that with the help of the
general relation (\ref{eq:general_estimate}) we can come back to estimating canonical
averages at any time during the multicanonical iteration. An alternative, rather
efficient, approach for arriving at a working estimate of $\Omega(E)$ was suggested
in Ref.~\cite{wang:01a}, where the weights $\omega(E)$ are {\em continuously\/}
updated $\omega(E) \rightarrow \omega(E) + \phi$ after visits of the energy $E$, and
the constant $\phi$ is gradually reduced to zero after the relevant energy range has
been sufficiently sampled. Although such a prescription ceases to form an equilibrium
Monte Carlo simulation, convergence to the correct density of states can be shown
under rather general circumstances \cite{zhou:05}.

Some combinations of the successful concepts of cluster algorithms/representations
and generalized-ensemble simulations have been suggested before, most notably the
multibondic algorithm of Ref.~\cite{wj:95a}, which attaches generalized weights to
the bond distribution function only (see also Ref.~\cite{hartmann:05}). Although it
appears most natural, it seems that it has not been noticed before that
multicanonical weights can be attached, instead, to the full random-cluster
probability density (\ref{eq:rc_probability}) to directly estimate the geometrical
density of states $g(b,n)$. In this ``multi-bondic-and-clusteric'' version one writes
\begin{equation}
  \label{eq:p_mubocl}
  p_\mathrm{mubocl}(b,n) = W_\mathrm{mubocl}^{-1}\,g(b,n)\,e^{-\gamma(b,n)},  
\end{equation}
such that the generalized weights $\exp[-\gamma(b,n)]$ lead to a completely flat
histogram for $\gamma(b,n) = \ln g(b,n)$. At this point, it is crucial to observe
that, since $g(b,n)$ is a purely combinatorial quantity describing the number of
decompositions of the lattice through a given number of activated links, it is no
longer necessary to simulate the underlying spin model and, instead, one can consider
the corresponding percolation problem directly. This approach proceeds by simulating
subgraphs ${\cal G}'$ with local updates: assume that the current subgraph consists
of $b$ active bonds resulting in a decomposition of the graph into $n$
clusters. Picking a bond of the graph ${\cal G}$ at random two local moves are
possible:
\begin{enumerate}
\item If the chosen bond is not active, try to activate it. Then either
  \begin{enumerate}
  \item activating the bond does not change the cluster number $n$ ({\em internal bond}),
    leading to a transition $(b,n) \rightarrow (b+1,n)$,
  \item or activating the bond does join two previously disjoint clusters ({\em
      coordinating bond\/}), such that $(b,n) \rightarrow (b+1,n-1)$.
  \end{enumerate}
\item If the chosen bond is already active, try to deactivate or delete it. Then
  either
  \begin{enumerate}
  \item deleting the bond does not change the cluster number $n$ (internal bond),
    resulting in the transition $(b,n) \rightarrow (b-1,n)$,
  \item or deleting the bond breaks a cluster apart in two parts (coordinating bond),
    such that $(b,n) \rightarrow (b-1,n+1)$.
  \end{enumerate}
\end{enumerate}
While with a naive approach (such as the application of the Hoshen-Kopelman algorithm
\cite{hoshen:79}) most of these moves would be very expensive computationally, this
is not the case for a clever choice of data structures and algorithms. We use
so-called ``union-find algorithms'' with additional improvements known as balanced
trees and path compression \cite{newman:01a}. With these structures, the
computational effort for identifying whether an inactive bond is internal or
coordinating and, for case (1b), the amalgamation of two clusters are operations in
constant running time, irrespective of the size of the graph (up to logarithmic
correction terms). The decision whether an active bond is internal or coordinating,
although an operation with ${\cal O}({\cal E})$ complexity in the worst case, can be
implemented very efficiently with interleaved breadth-first searches. Only the
operation (2b) of actually decomposing a cluster can be potentially expensive, but
this is only a problem directly at the percolation threshold. These local steps are
used for a generalized-ensemble simulation, for instance using the iteration
suggested by Wang and Landau \cite{wang:01a} to arrive at an estimate $\hat{g}(b,n)$
for the random-cluster density of states (additional speedups can be achieved
employing interpolation schemes for yet unvisited $(b,n)$ bins).

\begin{figure}[t]
  \centering
  \includegraphics[width=9cm]{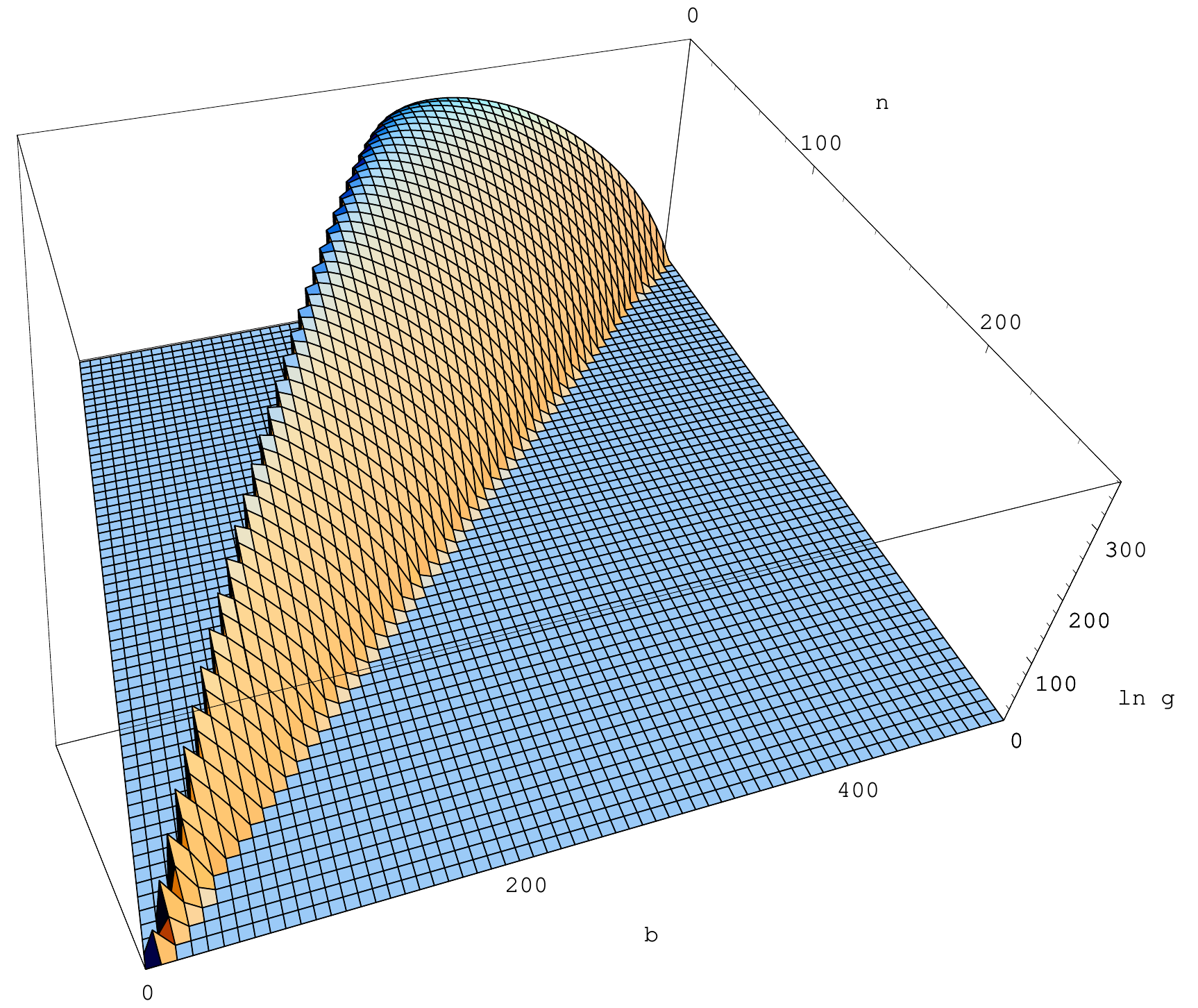}
  \vspace*{-0.5cm}
  \caption
  { Logarithm of the cluster density of states $g(b,n)$ of the $q$-state Potts model
    on a $16\times 16$ square lattice.
    \label{fig:density}
  }
\end{figure}

\begin{figure}[t]
  \centering
  \vspace*{0.3cm}
  \includegraphics[width=7.75cm]{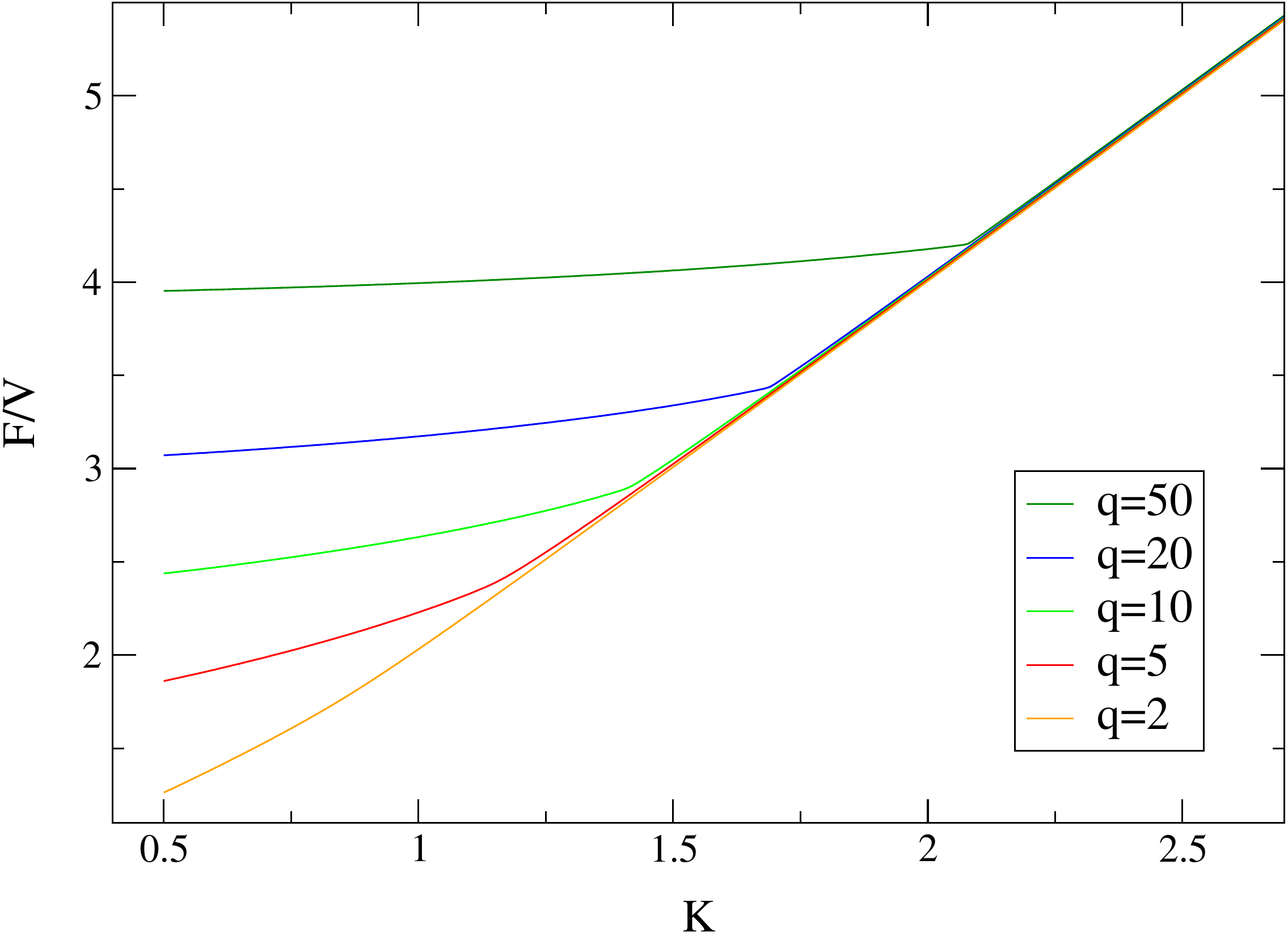}
  \includegraphics[width=7.75cm]{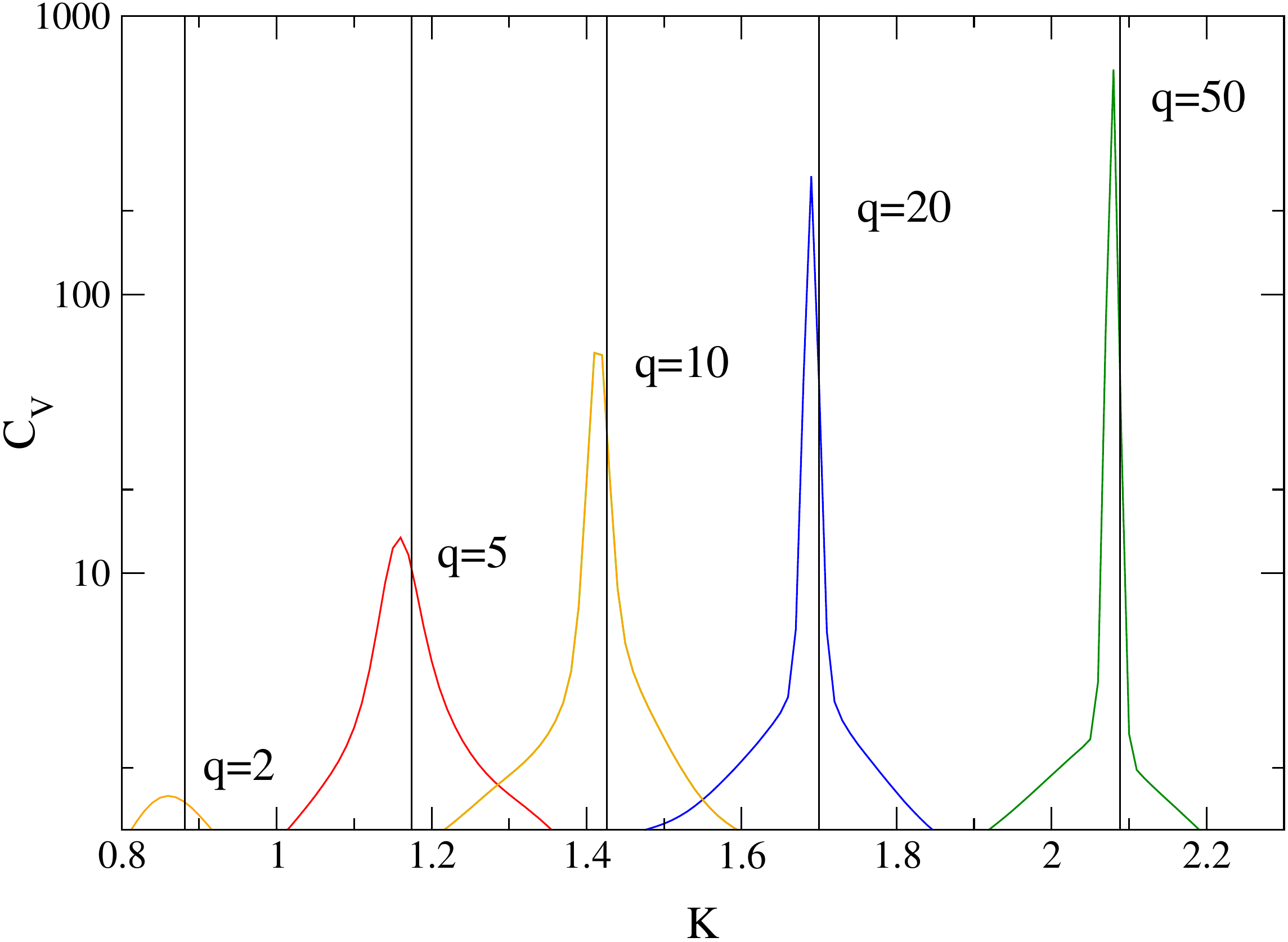}
  \vspace*{-0.5cm}
  \caption
  { Absolute free energy (left) and specific heat (right) of the $q=2$, $5$, $10$,
    $20$ and $50$ states Potts model on a $16\times 16$ square lattice as estimated
    from the density of states $\hat{g}(b,n)$ resulting from a
    ``multi-bondic-and-clusteric'' simulation described in the main text.
    \label{fig:potts}
  }
\end{figure}

The estimated $g(b,n)$ can then be used either for directly estimating thermal
averages via the relation (\ref{eq:partition_bcpm}) or as weight function for a
multi-bondic-and-clusteric simulation to yield estimates of arbitrary observables via
the general relation (\ref{eq:general_estimate}). Note that, by construction, the
approach does not suffer from any (hyper-)critical slowing down, since it is based
entirely on simulating a non-interacting percolation model. Figure \ref{fig:density}
shows the (logarithm of the) density of states $g(b,n)$ sampled with this approach on
a $16\times 16$ square lattice. While the $\hat{g}(b,n)$ resulting from this approach
still comes only up to an unknown normalization constant, the random-cluster approach
has the advantage that there exist ${\cal E}$ independent normalization conditions
\begin{equation}
  g(b) = \sum_{n} g(b,n) \stackrel{!}{=} {{\cal E} \choose b}.
\end{equation}
It is easily shown that the estimates from this approach reproduce the known results,
e.g., for the internal/free energy and specific heat \cite{ferdinand:69a} or the
(energy) density of states of the Ising model \cite{beale:96a}. Beyond that, it is
easy from this approach to study Potts model properties as a continuous function of
$q$, or to study equilibrium distributions of Potts models with a large number of
states without the problem of hypercritical slowing down. This is illustrated in
Figure \ref{fig:potts}.

\section{Conclusions}

While importance sampling Monte Carlo simulations according to the
Metropolis-Hastings scheme appear to be the universally optimal solution to the
problem of estimating equilibrium thermal averages, a number of complications are
encountered in practical applications which result (a) from the requirement of
computing estimates as continuous functions of external parameters and (b) the
Markovian nature of the algorithm entailing autocorrelations that can lead to dynamic
ergodicity breaking. I have outlined how a number of techniques such as histogram
reweighting, cluster algorithms and generalized-ensemble simulations can provide
(partial) solutions to these problems. It turns out that all of these techniques are
closely related to the problem of estimating the density of states of the model at
hand which turns out to be a central quantity for the understanding of advanced
simulation techniques. For the prototypic case of the Potts model, it is shown how a
combination of the random-cluster representation underlying the concept of cluster
algorithms and multicanonical simulations allows to reduce the simulation to a purely
geometric cluster counting problem that can be efficiently solved, e.g., with the
Wang-Landau sampling scheme to yield arbitrary thermal averages as continuous
functions of both the temperature and the (general, non-integer) number of states
$q$. Possible applications are investigations of the tricritical point in the $(T,q)$
plane, estimates of critical exponents as continuous functions of $q$, or the
investigation of transition states in the first-order regime, to name only a few of
the problems that immediately come to mind.

\section*{Acknowledgments}

The author acknowledges support by the DFG through the Emmy Noether Programme under
contract No.\ WE4425/1-1.


\end{document}